\documentclass[sigconf]{acmart}

\def\BibTeX{{\rm B\kern-.05em{\sc i\kern-.025em b}\kern-.08emT\kern-.1667em\lower.7ex\hbox{E}\kern-.125emX}}
    
\begin{document}

%
% The "title" command has an optional parameter, allowing the author to define a "short title" to be used in page headers.
\title[ECORS]{ECORS: An Ensembled Clustering Approach to Eradicate The Local And Global Outlier In Collaborative Filtering Recommender System}

%
% The "author" command and its associated commands are used to define the authors and their affiliations.
% Of note is the shared affiliation of the first two authors, and the "authornote" and "authornotemark" commands
% used to denote shared contribution to the research.
%\author{Surajit Das Barman}
%\email{surajitbarman012@ewubd.edu}
%\email{webmaster@marysville-ohio.com}
%\affiliation{%
  %\institution{Institute for Clarity in Documentation}
  %\streetaddress{P.O. Box 1212}
  %\city{Dhaka}
 % \state{Bangladesh}

%\author{Md. Asiful Kabir}
%\affiliation{%
%  \institution{Department of Computer Science and Engineering, East West University}
%  \streetaddress{East West University}
%  \city{Dhaka}
%  \country{Bangladesh}}
%\email{asif.heemel@gmail.com}

%\author{Aysha Akter Chamak}
%\affiliation{%
%  \institution{Department of Computer Science and Engineering, East West University}
%  \streetaddress{East West University}
%  \city{Dhaka}
%  \country{Bangladesh}}
%\email{ayshachamak@gmail.com}

\author{Mahamudul Hasan}
\affiliation{%
\institution{Department of Computer Science and Engineering,\\ University of Minnesota Twin Cities}
\country{Minneapolis, United States}}
\email{munna09bd@gmail.com}

%\author{Falguni Roy}
%\affiliation{%
 % \institution{Institute of Information Technology, Noakhali Science and Technology University}
%  \streetaddress{East West University}
%  \city{Noakhali}
%  \country{Bangladesh}}
%\email{falguniroy.iit@gmail.com}

%\affiliation{%
%  \institution{Tsinghua University}
%  \streetaddress{30 Shuangqing Rd}
% \city{Haidian Qu}

%\affiliation{%
%  \institution{Palmer Research Laboratories}
%  \streetaddress{8600 Datapoint Drive}

%\affiliation{\institution{The Th{\o}rv{\"a}ld Group}}
%\email{jsmith@affiliation.org}

%\affiliation{\institution{The Kumquat Consortium}}
%\email{jpkumquat@consortium.net}

%
% By default, the full list of authors will be used in the page headers. Often, this list is too long, and will overlap
% other information printed in the page headers. This command allows the author to define a more concise list
% of authors' names for this purpose.
%\renewcommand{\shortauthors}{Asif, et al.}

%
% The abstract is a short summary of the work to be presented in the article.
\begin{abstract}
Recommender systems are designed to suggest items based on user preferences, helping users navigate the vast amount of information available on the internet. Given the overwhelming content, outlier detection has emerged as a key research area in recommender systems. It involves identifying unusual or suspicious patterns in user behavior. However, existing studies in this field face several challenges, including the limited universality of algorithms, difficulties in selecting users, and a lack of optimization. In this paper, we propose an approach that addresses these challenges by employing various clustering algorithms. Specifically, we utilize a user-user matrix-based clustering technique to detect outliers. By constructing a user-user matrix, we can identify suspicious users in the system. Both local and global outliers are detected to ensure comprehensive analysis. Our experimental results demonstrate that this approach significantly improves the accuracy of outlier detection in recommender systems.

\end{abstract}

\keywords{Machine Learning, Recommender System, Information Search and Retrieval, Item-Item Based Collaborative Filtering, Cold-Start Problem}

%
% A "teaser" image appears between the author and affiliation information and the body 
% of the document, and typically spans the page. 
%  \Description{Enjoying the baseball game from the third-base seats. Ichiro Suzuki preparing to bat.}

%
% This command processes the author and affiliation and title information and builds
% the first part of the formatted document.
\maketitle

\section{Introduction}

The expeditious development of e-commerce and retail companies brings about a data overload problem. Recommendation systems have become productive and widely used to help people deal with data overload  \cite{a1}. In precise, recommender systems focus on predicting user preferences and try to recommend items that look to be interesting to them. The systems suggest user different items without having trouble. Recommender systems are broadly used in many websites like as videos (YouTube), movie recommendations (Netflix), product recommendations (Amazon), and music recommendations (Gnoosic) \cite{a2,a3}. Information needed for recommender systems is generally based on ratings and reviews from the user and purchase record. These ratings also can make the system vulnerable to attack. Many retailers try to upgrade the ratings of their products in recommendation items for their profit maximization. Various attackers inject fake profiles to affect the list. These attacks can be labeled as push attacks and nuke attacks. When attackers try to increase the rating of a targeted item, it's called a push attack and when attackers try to decrease the rating of a targeted item then it's called a nuke attack. Attackers influence the recommendation recurrence of specific items that they target by corrupting user profiles  \cite{a7}. Sometimes people give ratings randomly and this provides abnormal data. This fake profile and dummy ratings generate outliers in the dataset. An outlier is an inconsistent inspection that lies far away from remaining normal values in an arbitrary data sample. When a data point is far outside the entirety of the data, then it's called a global outlier on the other hand when data points are not outside the normal range but are abnormal depending on context and diverge significantly then it's a local outlier. Outlier detection has many paramount applications such as fraud detection, public health, measurement error detection, customer behavior analysis and so more. Collaborative filtering and content-based filtering are most popular for filtering information of the user. Collaborative filtering uses user interactions effectively to filter for those items that may be interesting to them. A compelling way of looking at collaborative filtering is that it is a concept of classification and regression. Content-based systems assemble recommendations using a user's content description and analyzing profile formation. Content-based recommendation systems analyze item descriptions to define items that are of particular preference to the user. Most of the Recommender systems generally use collaborative filtering algorithms to search users of similar preferences and choices \cite{a4,a5,a6}.

Clustering techniques can be used in a variant estate such as image processing, pattern recognition, statistical data exploration, and knowledge revelation \cite{a8}. Clustering algorithms may generate misleading partitions arising from local outliers in respect of different clusters. In this paper, we have used multiple clustering approaches to perceive the overall performance and difference of algorithms to find the global outlier and local outlier. Clustering-based approaches assert that outlier makes very small clusters while normal data objects belong to dense clusters  \cite{a9,a10}.

\begin{figure}[h]
  \centering
  \includegraphics[width=\linewidth]{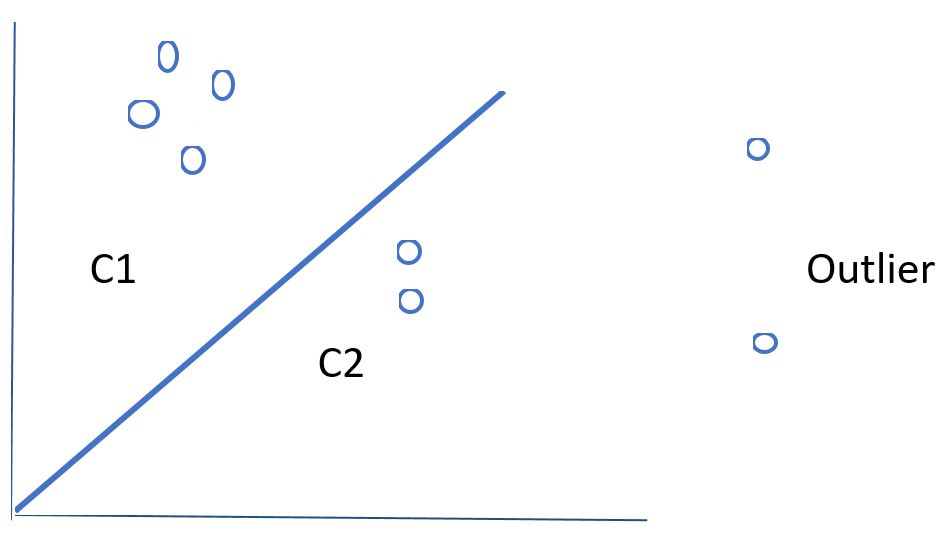}
  \caption{Outlier}
  %\Description{The 1907 Franklin Model D roadster.}
\end{figure}

\section{MOTIVATION}
  Recommender systems typically suggest top items of a recommended list. While the maximum item in a recommended list is very analogous it has become very comparable, so it expands the threat that the user may not choose one of these recommended items. Conversely, there is a broader possibility that the user may prefer at least one of these items when the recommended items are so different type. So when the user notices a completely contemporary item, the recommendation is precisely supportive. Replicated recommendations of popular items may decrease sales assortment. Desired books of a preferred list would scarcely be novel to the user. There are several approaches to get similarity but every way is not adequate to give an absolute result. As most of the data in real life are sparse that's why it should be handled wisely so that we can get a more exemplary result. If we work with more clustering algorithms and compare their results, we can get more canonical similarity between each cluster. Thus we can find which data is an outlier in its own clusters and also can predict global outlier specifically. So, we came up with an ensembled clustering approach to get a more unambiguous result. 

\section{BACKGROUND AND RELATED WORKS}
Recommender systems generally filter information to suggest items by predicting the ratings and choices of users. In the last several years, research in the domain of recommender systems has increased because of the rapid growth of e-commerce and the online world. Recommender systems are helpful to both retailers and users because of get more responses \cite{a11}. Based on different information categories, the Recommender system can be fabricated on rating and reviews of data, behavior pattern data like duration, transaction data like purchase extent and more. Many researchers proposed a different approach mainly based on content, collaborative filtering, item-user similarity, and hybrid recommendation   \cite{a12,a13,a14,a32}.
Outliers are obscure from other normal data point without adjacent data point in most manifestation. Most approaches for detecting outliers can be categorized into model-based, angle-based, proximity-based, similarity-based, and distribution-based approaches. In distribution-based approaches, outliers are predicted based on the probability distribution \cite{a15,a16}.  Researchers proposed some different clustering approaches \cite{a17,a18,a19,a34}. Lian Duan et al.\cite{a20} proposed a different approach to cluster-based outlier detection providing priority to the local data conduct and designing a clustering algorithm LDBSCAN to find clusters that have outliers. The clustering algorithm attempts to severance a group of data into different clusters to uncover cardinal groups within them.

Nilashi, M. et al.\cite{a35} had developed a recommender system for the tourism industry using SOM and EM for data clustering and HGPA for ensemble clustering tasks. Basically, the clustering algorithm divides the whole data set into several groups of data points to find significant groups and to compare the abnormal data in its cluster and data points in other clusters. Bobadilla, J. et al. \cite{a33} proposed a new pre-clustering algorithm using a method of a Bayesian non-negative matrix factorization (BNMF) to enhance clustering results in the collaborative filtering area. Clustering-based approaches fragment the points while density-based approaches fragment the space \cite{a23}.  Distance-based approaches apprise a data point 'p' as an outlier when it is smaller than 'q' points within the distance 'd' from 'p'. The values of 'q' and 'd' are handled by the user \cite{a21,a22}. Markos Markou et al. \cite{a24} proposed a new statistical approach in the domain of outlier detection. In density-based approaches, they evaluate the area density of the data point and predict those data points as outliers which are in the low dense domain  \cite{a25}. Many similarity measures are used \cite{a31} in collaborative filtering like PCC, CPCC, SPCC, Jaccard Similarity, MSD, JMSD, COS, and ACOS because of simple but sturdy performance.

\section{OUR METHODOLOGY}
We have proposed the user-user similarity metric based on user ratings. Then we propose a little modification to the existing prediction method. Due to the sparseness of the data, the collaborative filtering approach can't predict items with acceptable accuracy. Again, in most of the cases, they provide misleading results. This problem is known as a cold start problem. For our case, we calculate the similarity based on item ratings. 

\subsection{Proposed Method}
Mainly in this paper, we discuss our working procedure of outlier detection using different clustering approaches. This is how we use the algorithms. Here we have also discussed the Clustering method of the algorithms. We have discussed our main working part and then we have shown a similar result with sample data. Here we have tried to generate a sample similarity matrix that updates frequently after the implementation of each algorithm and finally how we get a structured user-user similarity matrix. Then to find the outlier we use the value that we got from the similarity matrix. Here we have visualized our contribution. First, we read the train data of ratings and splinted the user IDs, movie IDs, given ratings, and time stamps. Then we have formed a two-dimensional array (rat [u id] [mid]) for storing the rating that a user gave to a movie. We have also formed an array list (user Cluster) where we have stored movie IDs corresponding to each user ID to see which movies a user has rated.

Normalizing rating:
We Have normalized the ratings for calculation simplicity (0-1).

Distance/difference calculation:
We have considered the difference in ratings they gave to movies they both rated for determining the distance between two users.
Procedure:
We have compared user IDs in pairs and checked if there are any common movies that both users have rated. If there is, we calculated the difference between the ratings they have provided for each movie. We took the absolute value of the differences and summed them for all common movies. Then to find outliers we use the value to calculate the distance.

\begin{table}
  \caption{A user-item rating matrix}
  \label{tab:usermatrix}
  \begin{tabular}{ccccccc}
    \toprule
    Users & M1 & M2 & M3 & M4\\
    \midrule
    User1 		  &  5          &  3          &      -         &   4 \\ 
    User2 		  &  4          &  2          &      5         &   3 \\ 
    User3 		  &  -          &  4          &      5         &   2 \\
    User4 		  &  4          &  3          &      -         &   4 \\ 
    User5 		  &  4          &  2          &      5         &   3 \\ 
    User6 		  &  -          &  4          &      5         &   2 \\
    User7 		  &  -          &  -          &      -         &   - \\
    User8 		  &  -          &  -          &      -         &   - \\

  \bottomrule
\end{tabular}
\end{table}

Our experimental data table \ref{tab:usermatrix}, set with eight individual users and a total of four movies as items. From this dataset, we will calculate a user-user-based matrix which also can be recognized as the similarity between users. Missing values are denoted by the symbol -. First, we read the train data of ratings and splinted the user IDs, movie IDs, given ratings, and time stamps. Then we formed a two-dimensional array (rat [u id] [mid]) for storing the rating that a user gave to a movie. we also formed an array list (user Cluster) where we stored movie IDs corresponding to each user ID to see which movies a user has rated.

\subsection{User-User Similarity Matrix After Applying K-mean}
Tapas et al.\cite{a26}, \cite{a27} proposed a simple and efficient implementation of Lloyds k-mean clustering algorithm which is easy to implement. This requires only kd-tree which is the main data structure. They have built the practical workability in two ways\cite{a28}, \cite{a29}. The idea of a rule-based system is used to store and control or handle the knowledge to use information in an efficient way. They are usually used in artificial intelligence-based applications and many other types of research \cite{a30}.

If we apply this algorithm in a small space like our example matrix table \ref{tab:usermatrix}, then the result will be like table \ref{tab:K-meanmatrix}. After using the K-mean clustering approach we found the below cluster.

C1: u1, u2, u3, u4

C2: u5, u6

In Table \ref{tab:K-meanmatrix}, there are two clusters. Cluster C1 contains u1, u2, u3, u4 in the dataset. Cluster C2 contains u5, u6. Cluster C2 is tiny, containing just two users. Cluster C1 is large in comparison to C2. Therefore, a clustering-based method asserts that the two users in C2 are local outliers. u7 and u8 are the global outliers as they don't belong to any cluster. 
\begin{table}
  \caption{A user-user similarity matrix after applying K-mean}
  \label{tab:K-meanmatrix}
  \begin{tabular}{ccccccccc}
    \toprule
    Users & U1 & U2 & U3 & U4 & U5 & U6 & U7 & U8\\
    \midrule
    U1   &   1   &   1   &   1   &   1   &   -   &   -  &   -   &   -\\
    U2   &   1   &   1   &   1   &   1   &   -   &   -  &   -   &   -\\
    U3   &   1   &   1   &   1   &   1   &   -   &   -  &   -   &   -\\
    U4   &   1   &   1   &   1   &   1   &   -   &   -  &   -   &   -\\
    U5   &   -   &   -   &   -   &   -   &   1   &   1  &   -   &   -\\
    U6   &   -   &   -   &   -   &   -   &   1   &   1  &   -   &   -\\
    U7   &   -   &   -   &   -   &   -   &   -   &   -  &   -   &   -\\
    U8   &   -   &   -   &   -   &   -   &   -   &   -  &   -   &   -\\

  \bottomrule
\end{tabular}
\end{table}

\subsection{User-User Similarity Matrix After Applying K-medoids}
K.Shinde et al.\cite{k1} introduced a fast k-medoids clustering algorithm which is better than k-mean and k-medoids. Actually the paper proposed a mixed k-medoids cluster algorithm for better detection of outliers for the Recommender system. This clustering algorithm works in two phases \cite{k2}

If we apply this algorithm in a small space like our example matrix table \ref{tab:usermatrix}, then the result will be like table \ref{tab:K-medoidsmatrix}. After using the K-mediods clustering approach we found the below cluster.

C1: u1, u2, u3

C2: u4, u5, u6

In Table \ref{tab:K-medoidsmatrix}, there are two clusters. Cluster C1 contains u1, u2, u3 in the data set. Cluster C2 contains u4, u5, u6. Therefore, a clustering-based method asserts that the two users in C2 are local outliers for C1. On the other hand, u7 and u8 are global outliers as they don't belong to any cluster.

\begin{table}
  \caption{A user-user similarity matrix after applying K-Medoids}
  \label{tab:K-medoidsmatrix}
  \begin{tabular}{ccccccccc}
    \toprule
    Users & U1 & U2 & U3 & U4 & U5 & U6 & U7 & U8\\
    \midrule
    U1   &   2   &   2   &   2   &   1   &   -   &   -  &   -   &   -\\
    U2   &   2   &   2   &   2   &   1   &   -   &   -  &   -   &   -\\
    U3   &   2   &   2   &   2   &   1   &   -   &   -  &   -   &   -\\
    U4   &   1   &   1   &   1   &   2   &   1   &   1  &   -   &   -\\
    U5   &   -   &   -   &   -   &   1   &   2   &   2  &   -   &   -\\
    U6   &   -   &   -   &   -   &   1   &   2   &   2  &   -   &   -\\
    U7   &   -   &   -   &   -   &   -   &   -   &   -  &   -   &   -\\
    U8   &   -   &   -   &   -   &   -   &   -   &   -  &   -   &   -\\

  \bottomrule
\end{tabular}
\end{table}

\subsection{User-User Similarity Matrix After Applying DB-SCAN} Tsikrika et al.\cite{k3} worked with a combination of content and rating-based data. First, they find a similar neighbor and then they use DBSCAN to get the most similar users or items from that combination. DBSCAN clustering requires predefined epsilon value which defines an area and a minPts value which defines a minimum number of points required to be a cluster.

If we apply this algorithm in a small space like our example matrix table \ref{tab:usermatrix}, then the result will be like table \ref{tab:DBScanmatrix}. After using the DBSCAN clustering approach we found the below cluster.

cluster1: u1,u4,u5,u6

cluster2: u2,u3

In Table 2, there are two clusters. Cluster C1 contains u1, u4, u5,u6 in the data set. Cluster C2 contains u2, u3. Cluster C2 is tiny, containing just two users. Cluster C1 is large in comparison to C2. Therefore, a clustering-based method asserts that the two users in C2 are local outliers for C1. u7 and u8 are global outliers as they don't belong to any cluster.

\begin{table}
  \caption{A user-user similarity matrix after applying DB-SCAN}
  \label{tab:DBScanmatrix}
  \begin{tabular}{ccccccccc}
    \toprule
    Users & U1 & U2 & U3 & U4 & U5 & U6 & U7 & U8\\
    \midrule
    U1   &   3   &   2   &   2   &   2   &   1   &   1  &   -   &   -\\
    U2   &   2   &   3   &   3   &   1   &   -   &   -  &   -   &   -\\
    U3   &   2   &   3   &   3   &   1   &   -   &   -  &   -   &   -\\
    U4   &   2   &   1   &   1   &   3   &   2   &   2  &   -   &   -\\
    U5   &   1   &   -   &   -   &   2   &   3   &   3  &   -   &   -\\
    U6   &   -   &   1   &   1   &   1   &   2   &   2  &   -   &   -\\
    U7   &   -   &   -   &   -   &   -   &   -   &   -  &   -   &   -\\
    U8   &   -   &   -   &   -   &   -   &   -   &   -  &   -   &   -\\

  \bottomrule
\end{tabular}
\end{table}

\subsection{User-User Similarity Matrix After Applying Divisive}
If we apply this algorithm in a small space like our example matrix table \ref{tab:usermatrix}, then the result will be like table \ref{tab:Divisivematrix}. After using the Divisive clustering approach we found the below cluster.

cluster1: u1,u2,u3,u6

cluster2: u4,u5

In Table \ref{tab:Divisivematrix}, there are two clusters. Cluster C1 contains u1, u2, u3 u6. Cluster C2 contains u4 u5. So Cluster C2 is tiny, containing just two users. Cluster C1 is large in comparison to C2. Therefore, a clustering-based method asserts that the two users in C2 are local outliers. u7 and u8 are global outliers as they don't belong to any cluster.

\begin{table}
  \caption{A user-user similarity matrix after applying Divisive}
  \label{tab:Divisivematrix}
  \begin{tabular}{ccccccccc}
    \toprule
    Users & U1 & U2 & U3 & U4 & U5 & U6 & U7 & U8\\
    \midrule
    U1   &   4   &   3   &   3   &   2   &   1   &   2  &   -   &   -\\
    U2   &   3   &   4   &   4   &   1   &   -   &   1  &   -   &   -\\
    U3   &   3   &   4   &   4   &   1   &   -   &   1  &   -   &   -\\
    U4   &   2   &   1   &   1   &   4   &   3   &   2  &   -   &   -\\
    U5   &   1   &   -   &   -   &   3   &   4   &   3  &   -   &   -\\
    U6   &   2   &   1   &   1   &   2   &   3   &   4  &   -   &   -\\
    U7   &   -   &   -   &   -   &   -   &   -   &   -  &   -   &   -\\
    U8   &   -   &   -   &   -   &   -   &   -   &   -  &   -   &   -\\

  \bottomrule
\end{tabular}
\end{table}

\subsection{Flow Chart of Proposed Method}
Figure 2 shows the flowchart of our proposed method

\begin{figure}[h]
  \centering
  \includegraphics[width=\linewidth]{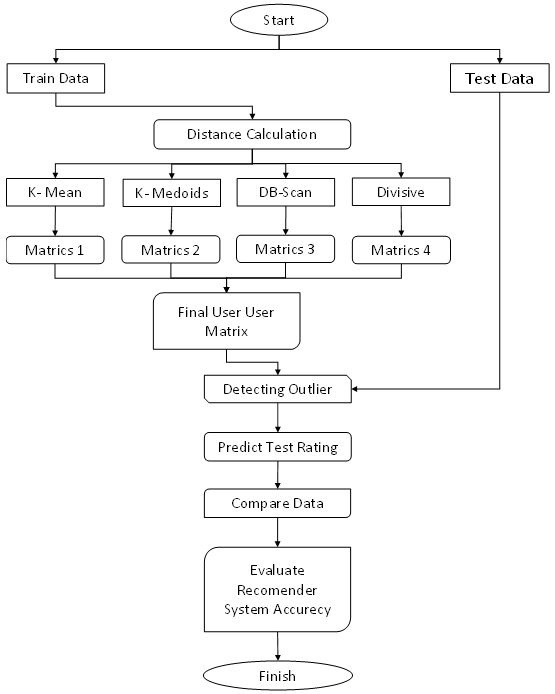}
  \caption{Flow Chart of Method}
  %\Description{The 1907 Franklin Model D roadster.}
\end{figure}

\section{EXPERIMENTAL SETUP}
\subsection{Dataset}
In this experiment, we used the Movielens dataset for our experimental purposes. MovieLens data set is called ML-1M, it includes 6040 users and 3952 movies with 1,000,209 ratings. The movie includes 19 genres.

\subsection{Metrics}
The most commonly used metrics include mean absolute error, precision, recall, and f-measure. Each metric has its strengths and weaknesses with respect to the task that is identified for the system.

\subsubsection{Mean Absolute Error}
Mean absolute error (MAE) has a place with a gathering of factual exactness measurements that think about the assessed appraisals against the genuine evaluations. All the more explicitly, MAE measures the normal outright deviation between an anticipated rating and the client's actual rating.

\begin{equation}
\label{MAE}
MAE= \frac{\sum_{i=1}^{N_u} |r_{u,m}- p_{u,m}|}{N_u}
\end{equation} 

However, proof proposes that different measurements indicate upgrades when the MAE esteem diminishes. This demonstrates MAE ought not to be limited as a potential measurement for topN suggestion errands.

\subsubsection{Precision}
Numerous specialists anticipate the rating of things by the client. The most utilized measurements are Mean Absolute Error (MAE) and Rooted Mean Squared Error (RMSE). Be that as it may, as a rule, best MAE or RMSE isn't equivalent to the best client fulfillment. The exactness and review are better in the topN proposal. Henceforth, so as to assess the execution of the proposed comparability display, the expectation exactness is estimated with two well-known utilized measurements.
The precision is computed as follows
:

\begin{equation}
\label{PRECISION}
Precision = \frac{n}{topN}
\end{equation} 

\subsubsection{Recall}
 The review score is the normal extent of things from the testing set that shows up among the positioned rundown from the preparation set. This measure ought to be as high as feasible for good execution. Accepting MT is the number of things that are in the testing set and loved by the dynamic client, and n is the measure of things that the testing client prefers and shows up in the prescribed rundown. 

Hence, the recall is computed as follows:

\begin{equation}
\label{RECALL}
Recall= \frac{N}{M_{T}}
\end{equation}  

\subsubsection{F-Measures}

In a factual examination of paired characterization, the F1 score (likewise F-score or F-measure) is a proportion of a test's precision. It considers both the accuracy p and the review r of the test to register the score: p is the quantity of right positive outcomes partitioned by the quantity of every positive outcome, and r is the quantity of right positive outcomes separated by the number of positive outcomes that ought to have been returned.

\begin{equation}
\label{f_measures}
F_{Measures} = \frac{2 \times Precision \times Recall}{Precision + Recall}
\end{equation}

\section{RESULTS AND DISCUSSIONS}
In this subsection, we actually discuss our result. Figure 2 to Figure 5 show the results gained by using the Movielens dataset. According to the figures, the results we obtained for all situations (MAE, precision, recall, and f-measures) with our proposed method are better than other traditional metrics.

\subsection{Performance Evaluation Using Movielens DataSet}
We used the Movielens dataset to evaluate our method. Figure 2 to Figure 5 shows the MAE, precision, recall, and f-measures obtained from the Movielens dataset. Using our outlier calculation of clustering approach by applying K-means, K-Medoids, DB-scan, and Divisive algorithms, we can detect outliers more effectively. In our procedure, we will get all users present in the similarity matrix.

Figure 2 to figure 5 shows the MAE, precision, recall, and f-measures obtained from Movielens data set when applying K-means, K-Medoids, DB-scan, and Divisive algorithms, and our proposed method named ECORS the user-user-based collaborative filtering method to detect outlier in a system.

%\begin{figure}[h]
%  \centering
%  \includegraphics[width=\linewidth]{1_4_1.pdf}
%  \caption{Top-N Neigbor vs. MAE \& Top-N Neighbor vs. Precision and Top-N Neigbor vs. Recall \& Top-N Neighbor vs. F-Measure Graph for Movielens Dataset.}
  %\Description{The 1907 Franklin Model D roadster.}
%\end{figure}

\begin{figure}[h]
  \centering
  \includegraphics[width=\linewidth]{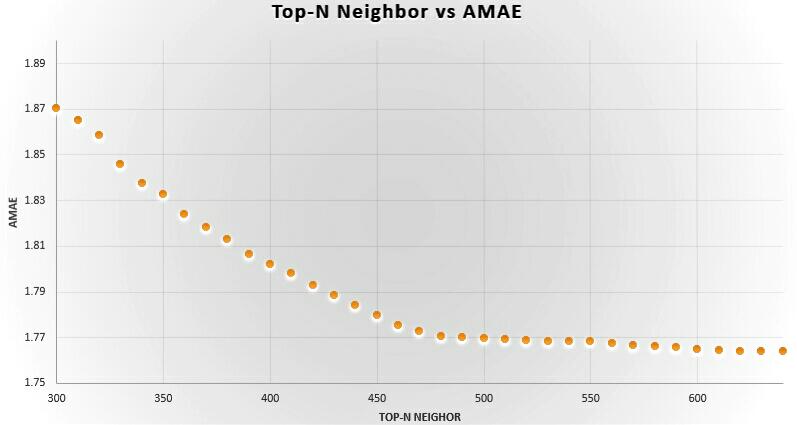}
  \caption{Top-N Neighbor vs. MAE.}
  %\Description{The 1907 Franklin Model D roadster.}
\end{figure}

\begin{figure}[h]
  \centering
  \includegraphics[width=\linewidth]{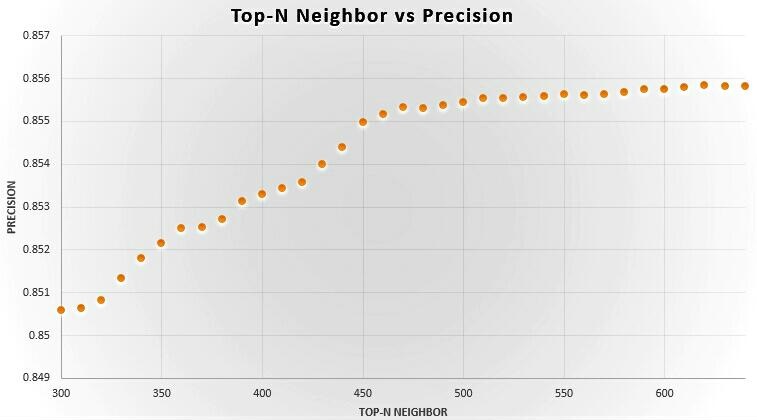}
  \caption{Top-N Neighbor vs. Precision.}
  %\Description{The 1907 Franklin Model D roadster.}
\end{figure}

\begin{figure}[h]
  \centering
  \includegraphics[width=\linewidth]{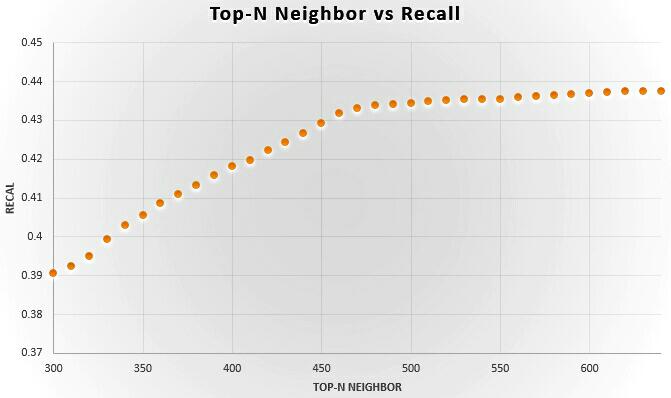}
  \caption{Top-N Neighbor vs. Recall.}
  %\Description{The 1907 Franklin Model D roadster.}
\end{figure}

\begin{figure}[h]
  \centering
  \includegraphics[width=\linewidth]{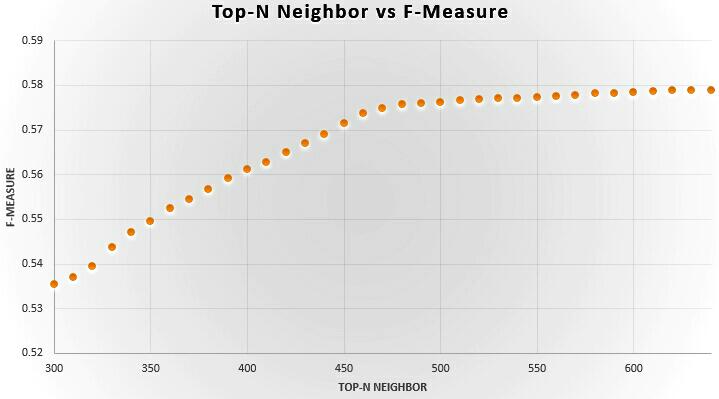}
  \caption{Top-N Neighbor vs. F-Measure.}
  %\Description{The 1907 Franklin Model D roadster.}
\end{figure}
%  \includegraphics[width=\textwidth]{1_4_1.pdf}
%  \caption{Top-N Neigbor vs. MAE \& Top-N Neighbor vs. Precision and Top-N Neigbor vs. Recall \& Top-N Neighbor vs. F-Measure Graph for Movielens Dataset.}
  %\Description{Enjoying the baseball game from the third-base seats. Ichiro Suzuki preparing to bat.}

\section{Conclusions}

In this paper, we first introduced a collaborative filtering method. Then we talked about the recommender system. We also discussed about user-based similarity matrix. We tried to give an overview of the cold-start problem in the recommender system and how the problem can be solved in a different way. Then we show our proposed method combining different types of clustering algorithms. We proposed an algorithm to generate a user-user similarity matrix to find the outlier between users.

%
% The next two lines define the bibliography style to be used, and the bibliography file.
\bibliographystyle{ACM-Reference-Format}
\bibliography{sample-base}

\end{document}